\documentclass[12pt,preprint]{aastex}

\usepackage{amsmath}

\newcommand{\CL}  {C.L.}
\newcommand{\dof} {d.o.f.}
\newcommand{\eVq} {\text{eV}^2}
\newcommand{\EtAl}{et al.}

\begin{document}


\title{Large mixing angle oscillations as a probe of the deep solar interior}
\author{C.~Burgess\altaffilmark{1},
  N.S.~Dzhalilov\altaffilmark{2},
  M.~Maltoni\altaffilmark{3},
  T.I.~Rashba\altaffilmark{2,3},
  V.B.~Semikoz\altaffilmark{2,3},
  M.A.~T\'ortola\altaffilmark{3} and
  J.W.F.~Valle\altaffilmark{3}}


\altaffiltext{1}{Physics Department, McGill University, 3600
  University Street; Montr\'eal, Qu\'ebec, Canada, H3A 2T8.}
\altaffiltext{2}{The Institute of Terrestrial Magnetism, Ionosphere
  and Radio Wave Propagation of the Russian Academy of Sciences,
  IZMIRAN, Troitsk, Moscow region, 142190, Russia.}
\altaffiltext{3}{Instituto de F\'{\i}sica Corpuscular --
  C.S.I.C./Universitat de Val{\`e}ncia, Edificio Institutos de
  Paterna, Apt 22085, E--46071 Val{\`e}ncia, Spain.,
  \texttt{http://alpha.ific.uv.es/\~{}ahep/}.}

\begin{abstract}
    We re-examine the sensitivity of solar neutrino oscillations to
    noise in the solar interior using the best current estimates of
    neutrino properties. Our results show that the measurement of
    neutrino properties at KamLAND provides new information about
    fluctuations in the solar environment on scales to which standard
    helioseismic constraints are largely insensitive. We also show how
    the determination of neutrino oscillation parameters from a
    combined fit of KamLAND and solar data depends strongly on the
    magnitude of solar density fluctuations.  We argue that a
    resonance between helioseismic and Alfv\'en waves might provide a
    physical mechanism for generating these fluctuations and, if so,
    neutrino-oscillation measurements could be used to constrain the
    size of magnetic fields deep within the solar radiative zone.
\end{abstract}

\keywords{elementary particles (neutrino) -- magnetohydrodynamics --
  sun: interior -- sun: oscillations -- sun: magnetic fields}

\maketitle

\section{Introduction}

Current solar \citep{ahm02a,ahm02b,Fukuda:2002pe,chlorine,sage,gallex}
and atmospheric \citep{Fukuda:1998mi} neutrino data give compelling
evidence that neutrino conversions take place. For the simplest case
of oscillations, the relevant parameters are two mass-squared
differences $\Delta m^2_{\mathrm{sol}}$ and $\Delta
m^2_{\mathrm{atm}}$, three angles $\theta_{12}$, $\theta_{23}$,
$\theta_{13}$ plus a number of CP violating
phases~\citep{Schechter:1980gr}.
One knows fairly well now that $\theta_{23}$ is nearly maximal (from
atmospheric data) and that the preferred solar solution for
$\theta_{12}$ is the so-called large mixing angle (LMA)
solution~\citep{Gonzalez-Garcia:2000aj}, while the third angle
$\theta_{13}$ is strongly constrained by the result of reactor
experiments~\citep{Apollonio:1999ae}. The CP phases are completely
unknown at present.
A recent analysis of solar and atmospheric data in terms of neutrino
oscillations is given in \citet{Maltoni:2002ni}, and finds the
currently-favored LMA solution of the solar neutrino problem has
\begin{equation} \label{eq:lmabfp}
    \tan^2\theta = 0.46\,, \quad 
    \Delta m^2 = 6.6\times 10^{-5}~\eVq
\end{equation}
and corresponds to oscillations into active neutrinos. Other recent 
analyses of solar data can be found in:
\citet{Fogli:2002pt,bah02,ban02a,ban02b,barger02,dho02,cre01}.

The recent results from the KamLAND reactor experiment~\citep{kamland}
have brought neutrino physics to a new stage. For the first time the
solar neutrino anomaly has been probed using terrestrial neutrino
sources.  This is fundamental for two reasons. First, among the
various proposed solutions of the solar neutrino problem, such as the
possibility of neutrino
spin-flavor-precession~\citep{Schechter:1981hw,akh88,lim88,mir01a,mir01b,bar02},
or non-standard neutrino matter interactions~\citep{Guzzo:2001mi},
which may arise in models of neutrino mass~\citep{moh86,hal86}, it
singles out a unique ``oscillation-type'' solution: the LMA MSW
solution.  Second, it brings to fruition one of the initial
motivations for studying solar neutrinos in the first
place~\citep{Bahcall}: the use of solar neutrinos to infer the
equilibrium properties of the solar core.

In this article we make the following points:
\begin{itemize}
  \item We show how the determination of neutrino oscillation
    parameters from a combined fit of KamLAND and solar data shows a
    strong dependence on the magnitude of solar density fluctuations.
    
  \item We show that the fact that the KamLAND results largely support
    LMA neutrino oscillations, can be used to provide new information
    about fluctuations in the solar core on much shorter scales than
    those which existing constraints (like helioseismology) can
    presently probe.
    
  \item We propose a physical process which could arise in the core,
    that may produce fluctuations on the scales to which solar
    neutrinos are sensitive.
\end{itemize}


\section{Effect of Fluctuations on Neutrino Propagation}

\begin{figure}[t]
    \plotone{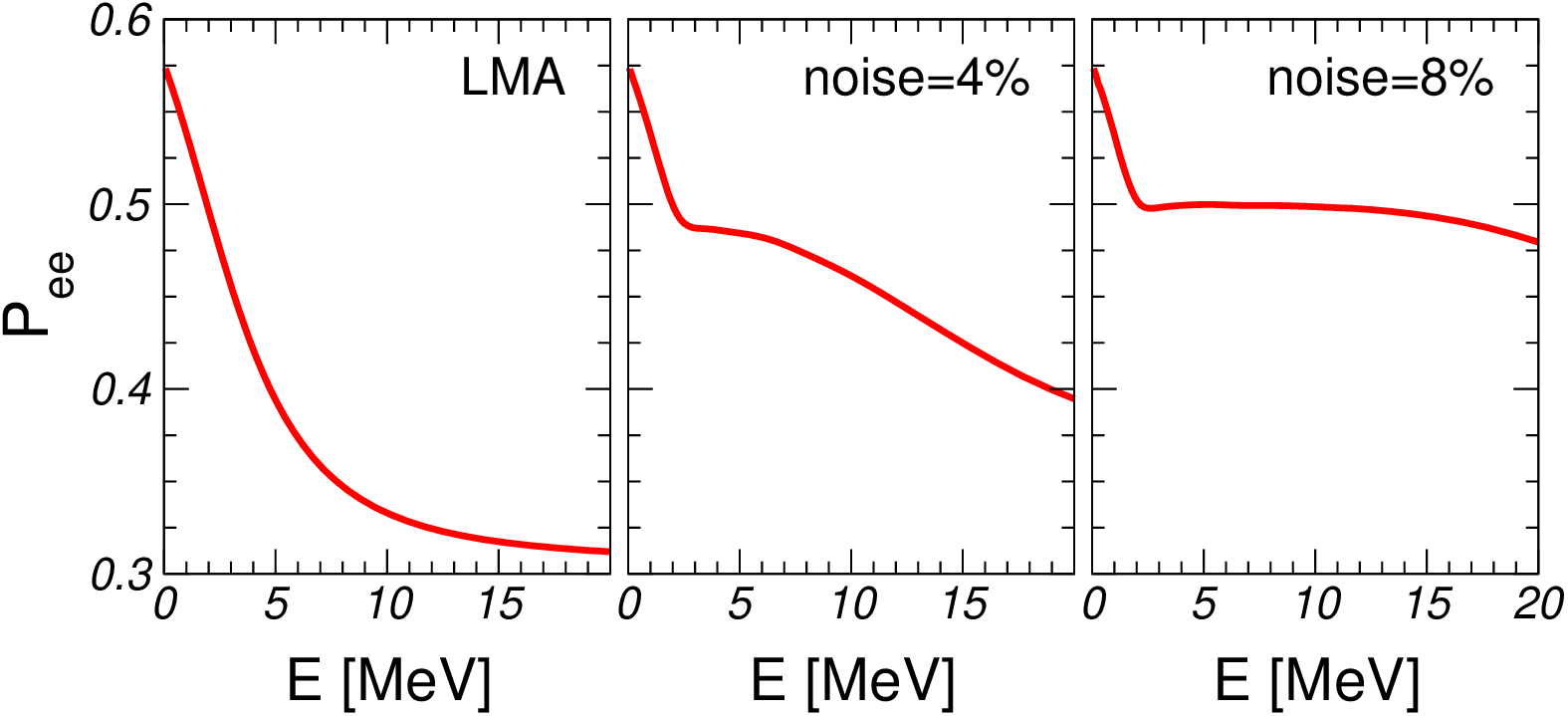}
    \caption{\label{fig:noisyLMA}%
      Survival probability of electron neutrinos for different levels
      of assumed random matter density perturbations in the LMA
      solution. The left panel is noiseless case, middle and right
      panels correspond to $\xi=4\%$ and $\xi=8\%$, respectively.}
\end{figure}

The evolution of solar neutrinos in the presence of a fluctuating
solar matter density has been considered previously by
\citet{Balantekin:1996pp,Bamert:1997jj,Nunokawa:1996qu,Burgess:1996}.
In a nutshell, these studies show that neutrino oscillations in the
Sun can be influenced by density fluctuations provided that two
conditions are satisfied {\em at the position of the MSW
oscillation}~\citep{Wolfenstein:1977ue,mik85}: ($i$) The fluctuation's
correlation length, $L_0$, is comparable to the local neutrino
oscillation length, $L_{\rm osc} \sim 100$~km and, ($ii$) the
fluctuation's amplitude, $\xi$, is at least roughly 1\%.  These
conclusions are summarized in Fig.~\ref{fig:noisyLMA}, where we show
how the electron-neutrino survival probability depends on the
fluctuation amplitude, $\xi$, given optimal choices for $L_0 = 100$~km
and neutrino oscillation parameters fixed at the best fit in
Eq.~\eqref{eq:lmabfp}.

These early estimates can now be sharpened in view of our better
understanding of neutrino-oscillation parameters. To illustrate this
we have performed a global analysis of the solar data, including
radiochemical experiments (Chlorine, Gallex-GNO and SAGE) as well as
the latest SNO data in the form of 17 (day) + 17 (night) recoil energy
bins (which include CC, ES and NC contributions,
see~\citet{Maltoni:2002ni})~\citep{ahm02a,ahm02b}, the
Super-Kamiokande spectra in the form of 44 bins~\citep{Fukuda:2002pe}
and the KamLAND rates and spectra given in \citet{kamland}.

The combined analysis of KamLAND and solar neutrino data with a
fluctuating solar medium depends on four parameters: the two neutrino
oscillation parameters $\Delta m^2$ and $\tan^2\theta$ and the two
parameters $\xi$ and $L_0 $ characterizing solar noise. In what
follows we present the results of two different analyses. First we
display the allowed neutrino oscillation parameters for given
assumption about the solar noise parameters. Later we present the
limits on solar noise parameters for given assumptions about the solar
neutrino oscillation parameters.
  
\section{Constraints on Solar Noise}

\begin{figure}[t]
    \plotone{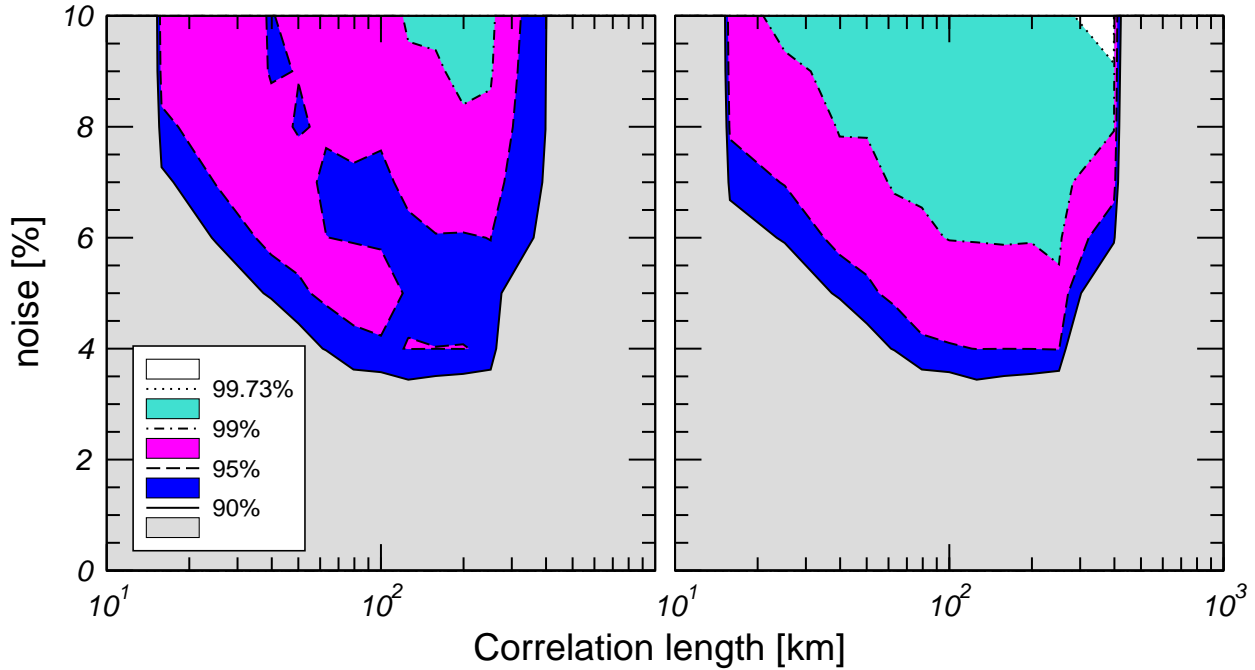}
    \caption{\label{fig:chi2fit}%
      Allowed regions for the solar noise parameters $L_0$ and $\xi$
      from the analysis of present solar neutrino~+~KamLAND data, when
      neutrino oscillation parameters are varied freely (left panel),
      or fixed at the present LMA best fit point (right panel).}
\end{figure}

The sensitivity of the neutrino signal to the solar density
fluctuations is shown in Fig.~\ref{fig:chi2fit}.  The regions denote
the bounds on $\xi$ as a function of the correlation length $L_0$ at
different confidence levels for 2 \dof.

Taking into account only the current solar neutrino data as in
\citep{Maltoni:2002ni} with the neutrino oscillation parameters in the
region $\Delta m^2 = 10^{-6}\div 10^{-3}~\eVq$ and $\tan^2\theta =
0.1\div 1$, we have found a very weak bound on the noise.  For
$L_0<100$~km or large $L_0>300$~km there is essentially no constraint.
Only for intermediate $L_0$ values inside the region $100$~km
$<L_0<300$~km we find that $\xi<8\%$ at 90\% \CL.

In contrast, the inclusion of the KamLAND reactor data constrains the
level of the solar density noise, irrespective of the values of
neutrino oscillation parameters in the above region.  The increased
sensitivity on the solar density noise parameters found in the global
analyses of solar + KamLAND data is seen in the left panel of
Fig.~\ref{fig:chi2fit}. If the solar neutrino parameters were known
with higher accuracy one could use them to probe the solar noise level
with even better sensitivity.  For example, fixing the neutrino
oscillation parameters at the current global solar + KamLAND best-fit
LMA point given in \citet{Maltoni:2002aw}, yields the bound given in
the right panel of Fig.~\ref{fig:chi2fit}.

\begin{figure}[t]
    \plotone{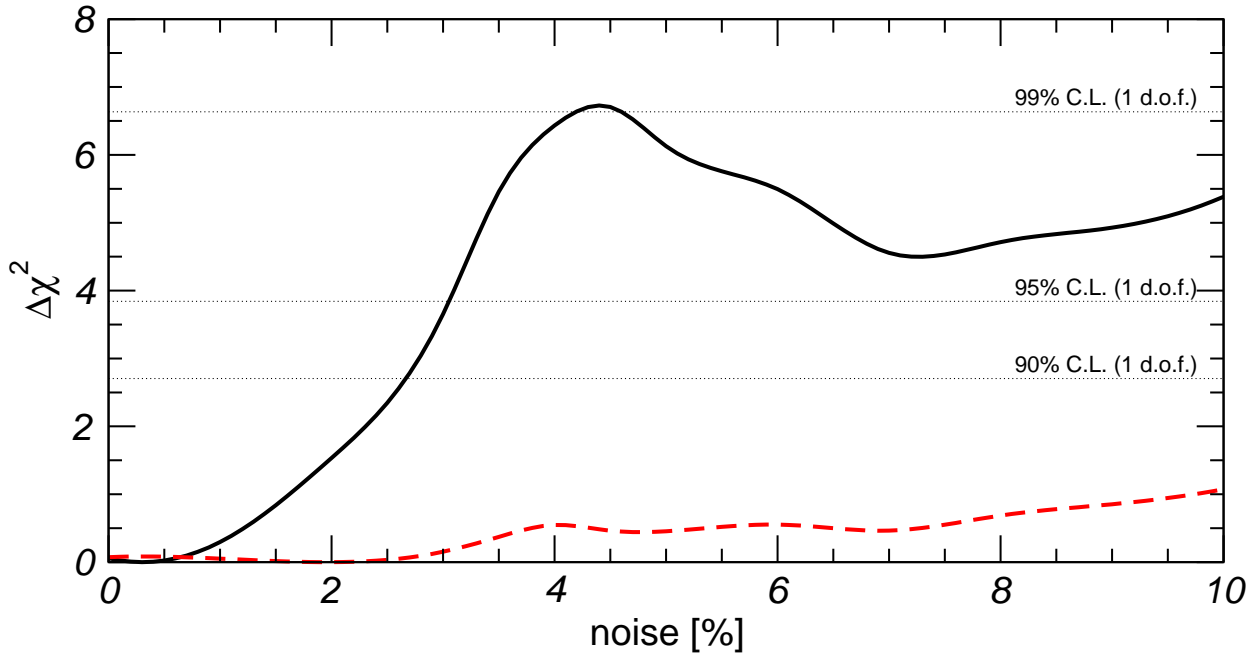}
    \caption{\label{fig.noise-chisq.eps}%
      $\chi^2$ versus noise strength for $L_0=100$~km after (upper)
      and before (lower curve) KamLAND.}
\end{figure}

Another way to present the global analysis of solar+KamLAND results is
illustrated in Fig.~\ref{fig.noise-chisq.eps}.
Here we display the minimal $\chi^2$ versus noise amplitude before and
after KamLAND, irrespective of neutrino parameters.
All in all, for a density fluctuation scale $L_0$ of 100~km or so, one
finds from this figure that the current limit to the noise amplitude
is about 3\% at 95\% \CL.  Note that these bounds follow from the
neutrino data themselves, no helioseismological argument can at
present rule out their existence.

\section{Effect of Fluctuations on Neutrino Parameter Determination }

\begin{figure}[t]
    \plotone{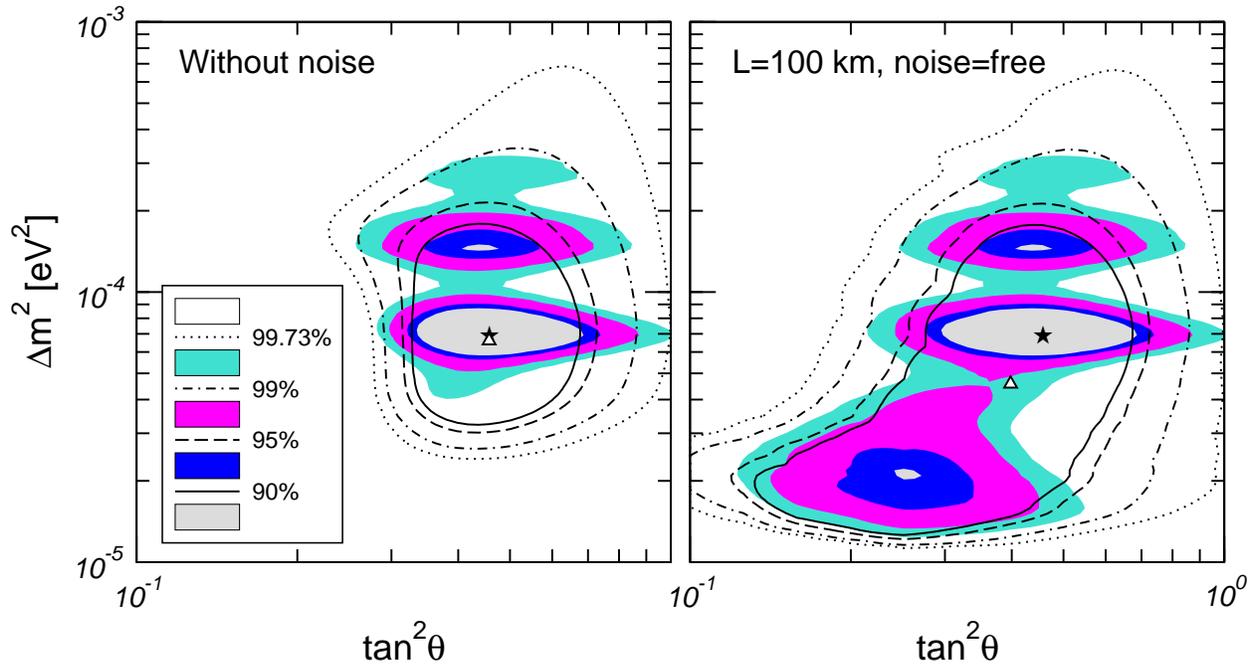}
    \caption{\label{fig.osc-region.eps}%
      Allowed regions for neutrino oscillation parameters $\Delta m^2$
      and $\theta$, for the ``standard'' noiseless Sun (left panel)
      and for a noisy Sun with arbitrary density noise magnitude $\xi$
      on a $L_0=100$~km spatial scale.  The lines and shaded regions
      correspond to the analyses of solar and solar+KamLAND data
      respectively. The best-fit point is indicated by a triangle
      (solar) or a star (solar+KamLAND).}
\end{figure}

Conversely, for given assumptions on the solar noise parameters, the
combined solar and KamLAND data can be used to determine the allowed
region of neutrino mixing parameters, as shown in
Fig.~\ref{fig.osc-region.eps}. First we give, in the left panel, the
allowed LMA-MSW region for a smooth solar density profile (noiseless
case), from \cite{Maltoni:2002ni}. The effect of the noise is shown in
the right panel, where we have fixed the spatial scale of fluctuation
at its optimal value $L_0=100$~km and left the noise magnitude free.
One sees that, taking into account solar neutrino data alone the
allowed region of neutrino oscillation parameters becomes bigger.  As
seen from the right panel, the inclusion of KamLAND data implies the
existence of a new region with substantially lower $\Delta m^2$
values, around $\Delta m^2=2 \times 10^{-5}~\eVq$, and somewhat lower
$\tan^2\theta$ around $\tan^2\theta=0.25$.  This new region is present
even at 90\% \CL\ with 2 \dof. One finds that the global best fit
point for neutrino parameters at $\xi=8\%$ noise level and $L_0=100$
km becomes $\Delta m^2=2 \times 10^{-5}~\eVq$ and $\tan^2\theta=0.25$,
a region excluded by the standard KamLAND + solar neutrino data
analysis without noise.  This illustrates how a precise knowledge of
the solar interior (the solar noise level) is required in order to
sharpen the determination of neutrino oscillation parameters.

Two objections were believed to limit this kind of analysis.  First,
on the observational side, the success of helioseismology seemed to
preclude the existence of fluctuation amplitudes larger than 1\% in
size. Second, on the theoretical side, no known physics of the solar
core could generate fluctuations large enough to be detected. Of these
two, the first is the more serious, since our inability to guess a
source of fluctuations in such a complicated environment is much less
worrying than is a potential conflict with helioseismic data.
Nonetheless, in the remainder of this letter we argue why neither of
these objections can rule out the possibility of having large enough
density fluctuations without undergoing more careful scrutiny.


\section{Helioseismic Bounds and Fluctuation Mechanism}

Helioseismology~\citep{Castellani:1997pk,Christensen-Dalsgaard:2002ur}
is rightfully celebrated as a precision tool for studying the inner
properties of the Sun. Careful measurements have provided precise
frequencies for numerous oscillation modes, and these may be compared
with calculations of these frequencies given assumed density and
temperature profiles for the solar interior. Constraints on solar
properties arise because careful comparison between theory and
measurements gives agreement only if the assumed profiles are within
roughly 1\% of the predictions of the best solar models.

For the present purposes, the weak link in this train of argument lies
in the details of the inversion process which obtains the solar
density profile given an observed spectrum of helioseismic
frequencies. This inversion is only possible if certain smoothness
assumptions are made about solar properties, due to the inevitable
uncertainties which arise in the observed solar helioseismic
oscillation pattern.  As a result helioseismology severely constrains
the existence of density fluctuations, but only those which vary over
very long scales $\gg$ 1000~km
\citep{Castellani:1997pk,Christensen-Dalsgaard:2002ur}.  In
particular, the measured spectrum of helioseismic waves is largely
insensitive to the existence of density variations whose wavelength is
short enough -- on scales close to $L_{\rm osc} \sim 100$~km, deep
within the solar core -- to be of interest for neutrino oscillations.
In particular, we claim that such density variations with amplitudes
as large as 10\% cannot yet be ruled out by helioseismic data.

\subsubsection*{Fluctuation Mechanism}

A mechanism which might produce density variations of the required
amplitude and correlation length arises once helioseismology is
reconsidered in the presence of magnetic fields, which are normally
neglected in helioseismic analyses \citep{Couvidat:2002bs}.
Generally, the neglect of magnetic fields in helioseismology is very
reasonable since the expected magnetic field energy densities,
$B^2/8\pi$, are much smaller than are gas pressures and other relevant
energies.

In \citep{newhelio} we study helioseismic waves in the Sun, and find,
contrary to naive expectations, that reasonable magnetic fields in the
radiative zone \citep{boruta,Parker} can appreciably affect the
profiles of helioseismic $g$-modes as a function of solar depth. In
particular, density profiles due to these waves tend to form spikes at
specific radii within the Sun, corresponding to radii where the
frequencies of magnetic Alfv\'en modes cross those of buoyancy-driven
($g$-type) gravity modes. Due to this resonance, energy initially in
$g$-modes is directly pumped into the Alfv\'en waves, causing an
amplification of the density profiles in the vicinity of the resonant
radius. This amplification continues until it is balanced by
dissipation, resulting in an unexpectedly large density variation at
the resonant radii. Furthermore, these level crossings only occur with
$g$-modes, and so typically occur deep within the solar radiative
zone. They also do not affect substantially the observed $p$-modes,
which makes it unlikely that these resonances alter standard analyses
of helioseismic data (which ignore solar magnetic fields), in any
significant way.

\begin{figure}[t]
    \plotone{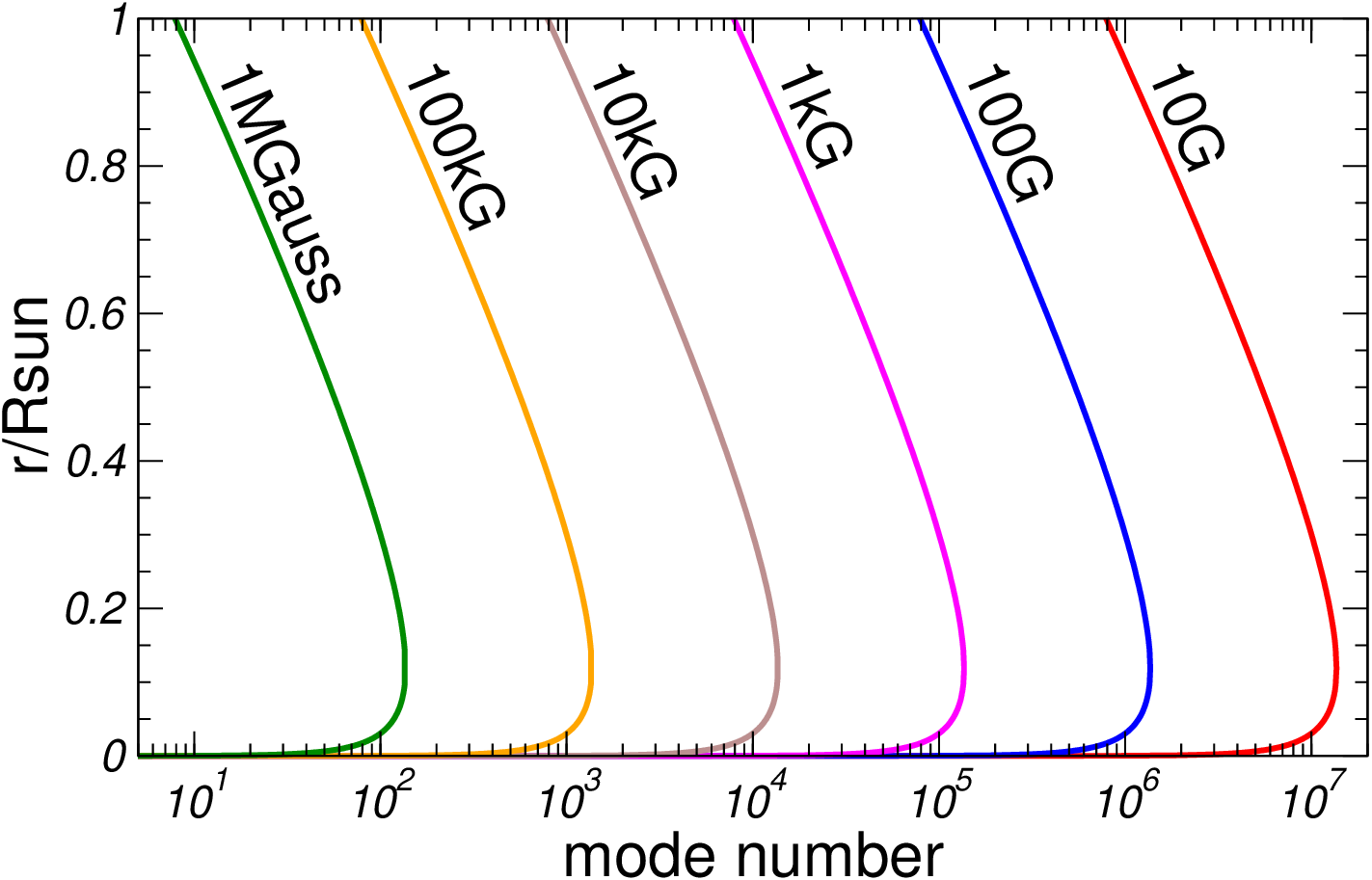}
    \caption{\label{fig:zepsR}%
      The positions of Alfv\'en/$g$-mode resonances as a function of
      mode number, for various magnetic field strengths.}
\end{figure}

Fig.~\ref{fig:zepsR} plots the positions of the Alfv\'en/$g$-mode
resonant layers as a function of an integer mode label $n$, for
various values of a hypothetical magnetic field. What this figure
shows is that there are very many such level crossings, whose position
varies most quickly with radius within the Sun near the solar center.
The superposition of several different modes results in a series of
relatively sharp spikes in the radial density profile at the radii
where these resonances take place. From the point of view of exiting
neutrinos, passage through these successive helioseismic resonances
mimics the passage through a noisy environment whose correlation
length is the spacing between the density spikes.

\begin{figure}[t]
    \plotone{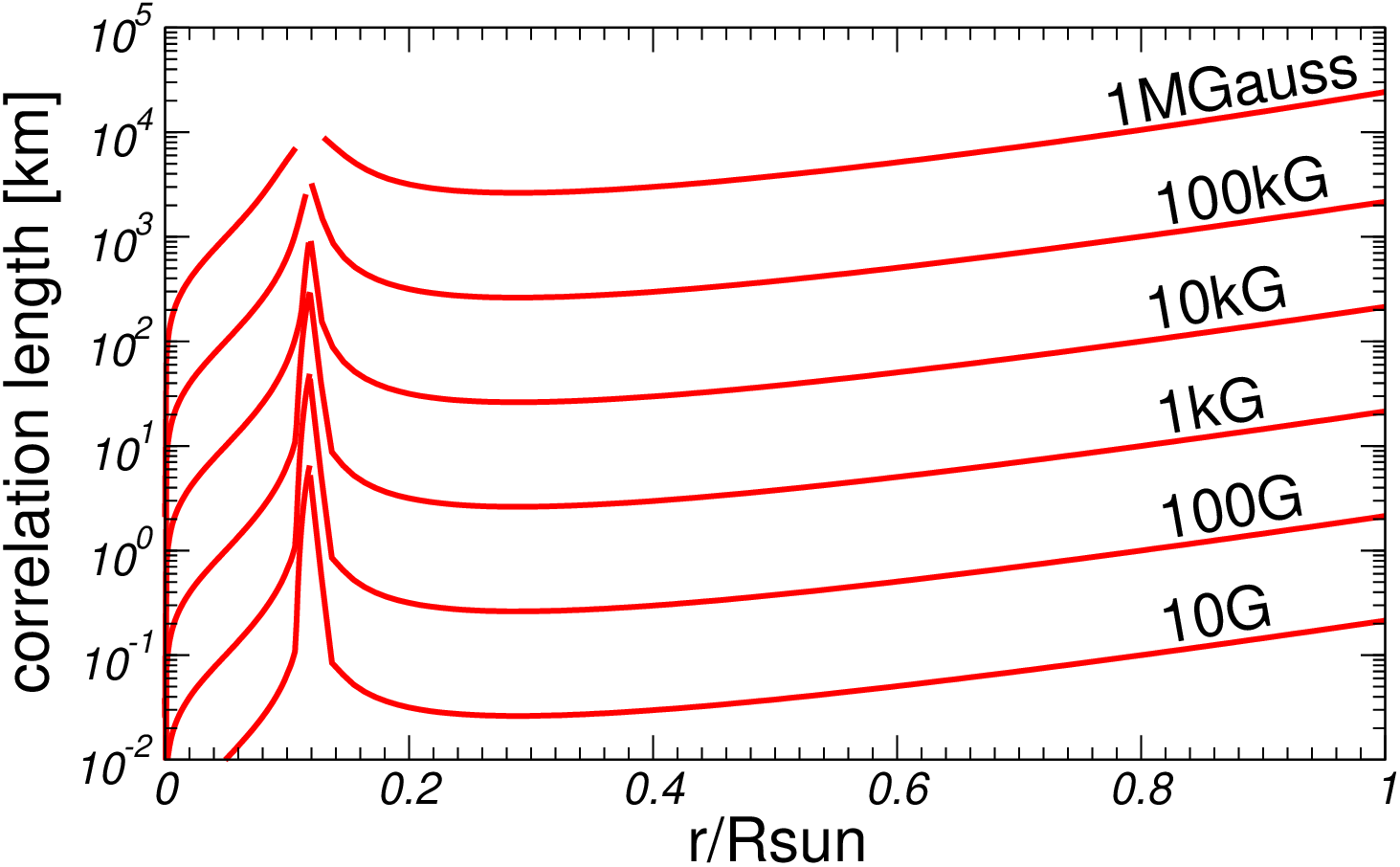}
    \caption{\label{fig:corr}%
      Solar density correlation length, versus distance from solar
      center, for different magnetic fields.}
\end{figure}

In \citep{newhelio} we make several estimates of the size of these
waves in the Sun. Fig.~\ref{fig:corr} shows how the spacing between
the spikes varies as a function of their position within the Sun. Note
that, while typically the width of the spikes is comparable to their
separation, and therefore causes important effects in neutrino
propagation, it is however considerably narrower than characteristic
g-waves.  As a result we find that the energy cost of producing them
with amplitudes as large as 10\% can be much less than the
prohibitively large values which were required to obtain similar
amplitudes for g-waves alone~\citep{Bamert:1997jj}.  It is remarkable
that the position at which this spacing is close to 100~km lies near
$r = 0.12 \; R_\odot$ for a very wide range of magnetic fields. On the
other hand, we have observed~\citep{newhelio} that, for magnetic
fields which are of order 10~kG, the spacing of resonances is near
100~km for a very wide range of radii -- including the neutrino
resonance region, $r \sim 0.3 \; R_\odot$.

Are such large radiative-zone magnetic fields possible? Very little is
directly known about magnetic field strengths within the radiative
zone. The only generally-applicable bound there is due to
Chandrasekar, and states that the magnetic field energy must be less
than the gravitational binding energy: $B^2/8\pi <
GM^2_\odot/R_\odot^4$, or $B < 10^8$ G. A stronger bound is also
possible if one assumes the solar magnetic field to be a relic of the
primordial field of the collapsing gas cloud from which the Sun
formed. In this case it has been argued that central fields cannot
exceed around 30 G \citep{boruta}. (Still stronger limits, $B <
10^{-3}$ G, are possible \citep{MestelWeiss} if the solar core should
be rapidly rotating, as is sometimes proposed.) Since the initial
origin of the central magnetic field is unclear, we believe any
magnetic field up to the Chandrasekar bound should be entertained.

The above mechanism has many of the features required to produce
density fluctuations in the Sun which may be relevant to the analysis
of solar neutrino data. Although much more study is required to
establish whether these resonances really occur and affect neutrinos,
we believe their potential existence substantially reinforces the
general motivation for using neutrino oscillations to directly probe
short-wavelength density fluctuations deep within the solar core.

\section{Summary and conclusion}
\label{sec:summary-conclusion}

We have re-examined the sensitivity of solar neutrino oscillations to
fluctuations in the solar density profile, using the best current
estimates of neutrino properties, especially the new reactor data from
KamLAND. Our results show that the measurement of neutrino properties
in the latter experiment provides new information about fluctuations
in the solar environment on scales to which standard helioseismic
constraints are largely insensitive. Conversely we have seen how the
determination of solar neutrino parameters from a fit of the data in
the case of a noisy Sun differ from the quiet Sun case.  We have also
argued that a resonance between helioseismic and Alfv\'en waves can
provide a physical origin for such fluctuations and, if so,
neutrino-oscillation measurements could be used to constrain the size
of magnetic fields deep within the solar radiative zone.

\acknowledgments

This work was supported by Spanish grants BFM2002-00345, by the
European Commission RTN network HPRN-CT-2000-00148, by the European
Science Foundation network grant N.~86, by Iberdrola Foundation (VBS)
and by INTAS grant YSF 2001/2-148 and MECD grant SB2000-0464 (TIR).
C.B.'s research is supported by grants from NSERC (Canada) and FCAR
(Quebec).  M.M.\ is supported by contract HPMF-CT-2000-01008.  VBS,
NSD and TIR were partially supported by the RFBR grant 00-02-16271.
M.A.T.\ was supported by the MECD fellowship AP2000-1953.


\begin{thebibliography}{}
  
\bibitem[Abdurashitov \EtAl(2002)]{sage} 
  Abdurashitov, D.~N., \EtAl\ 2002, preprint (astro-ph/0204245)
  
\bibitem[Ahmad \EtAl(2002a)]{ahm02a} Ahmad, Q.~R., \EtAl\ 2002a, 
  Phys.\ Rev.\ Lett.,\  89, 011301
  
\bibitem[Ahmad \EtAl(2002b)]{ahm02b} Ahmad, Q.~R., \EtAl\ 2002b,
  Phys.\ Rev.\ Lett.,\  89, 011302 
  
\bibitem[Akhmedov(1988)]{akh88}
  Akhmedov, E.~K. 1988,
  Phys.\ Lett.\ B, 213, 64
  
\bibitem[Apollonio \EtAl(1999)]{Apollonio:1999ae}
  Apollonio, M., \EtAl\ 1999, 
  Phys.\ Lett.\ B, 466, 415 
  
\bibitem[Bahcall(1989)]{Bahcall} 
  Bahcall, J. 1989, Neutrino Astrophysics,
  Cambridge University Press
  
\bibitem[Bahcall, Gonzalez-Garcia \& Pena-Garay(2002)]{bah02}
  Bahcall, J.~N., Gonzalez-Garcia, M.~C., \& Pena-Garay, C. 2002,
  JHEP, 0207, 054 
  
\bibitem[Balantekin, Fetter, \& Loreti(1996)]{Balantekin:1996pp} 
  Balantekin, A.~B., Fetter, J.~M., \& Loreti, F.~N. 1996,
  Phys.\ Rev.\ D, 54, 3941
  
\bibitem[Bamert, Burgess, \& Michaud(1997)]{Bamert:1997jj}
  Bamert, P., Burgess, C.~P.,  \& Michaud, D. 1997
  Nucl.\ Phys.\ B, 513, 319 
  
\bibitem[Bandyopadhyay \EtAl(2002a)]{ban02a}
  Bandyopadhyay, A., Choubey, S., Goswami, S., \& Roy, D.~P. 2002a,
  Phys.\ Lett.\ B, 540, 14 
  
\bibitem[Bandyopadhyay \EtAl(2002b)]{ban02b}
  Bandyopadhyay, A., Choubey, S., Goswami, S., \& Roy, D.~P. 2002b,
  Mod.\ Phys.\ Lett.\ A, 17, 1455
  
\bibitem[Barger \EtAl(2002)]{barger02}
  Barger, V., \EtAl, 2002,
  Phys.\ Lett.\ B, 537, 179
  
\bibitem[Barranco \EtAl(2002)]{bar02}
  Barranco, J., \EtAl\ 2002,
  Phys.\ Rev.\ D, 66, 93009
  
\bibitem[Boruta(1996)]{boruta} 
  Boruta, N. 1996, ApJ, 458, 832
  
\bibitem[Burgess \& Michaud(1996)]{Burgess:1996}
  Burgess, C.~P., \& Michaud, D. 1996
  Ann. Phys. (NY), 256, 1
  
\bibitem[Burgess \EtAl(2003)]{newhelio} 
  Burgess, C.~P., Dzhalilov, N.~S., Rashba, T.~I., Semikoz, V.~B.,
  \& Valle, J.~W.~F. 2003, in preparation
  
\bibitem[Castellani \EtAl(1997)]{Castellani:1997pk}
  Castellani, V., \EtAl\ 1997,
  Nucl.\ Phys.\ Proc.\ Suppl.,\ 70, 301
  
\bibitem[Cleveland \EtAl(1998)]{chlorine} 
  Cleveland, B.~T., \EtAl\ 1998,
  ApJ, 496, 505 
  
\bibitem[Christensen-Dalsgaard(2002)] {Christensen-Dalsgaard:2002ur} 
  Christensen-Dalsgaard, J. 2002, preprint (astro-ph/0207403).
  Lecture Notes available at
  \verb"http://bigcat.obs.aau.dk/~jcd/oscilnotes/"
  
\bibitem[Couvidat, Turck-Chieze, \& Kosovichev(2002)]{Couvidat:2002bs}
  Couvidat, S., Turck-Chieze, S., \& Kosovichev, A.~G. 2002,
  preprint (astro-ph/0203107)
  
\bibitem[Creminelli, Signorelli \& Strumia(2001)]{cre01}
  Creminelli, P., Signorelli, G., \& Strumia, A. 2001,
  JHEP, 0105, 05 2
  
\bibitem[de Holanda \& Smirnov(2002)]{dho02}
  de Holanda, P.~C., \& Smirnov, A.~Y. 2002,
  preprint (hep-ph/0205241)
  
\bibitem[Eguchi \EtAl(2002)]{kamland}  
  Eguchi, K.\ \EtAl\ [KamLAND Collaboration] 2003,
  Phys.\ Rev.\ Lett., 90, 021802 
  
\bibitem[Fogli \EtAl(2002)]{Fogli:2002pt}
  Fogli, G.~L., \EtAl, 2002
  Phys.\ Rev.\ D, 66, 053010
  
\bibitem[Fukuda \EtAl(2002)]{Fukuda:2002pe} Fukuda, S., \EtAl\ 2002, 
  Phys.\ Lett.\ B, 539, 179
  
\bibitem[Fukuda \EtAl(1998)]{Fukuda:1998mi} Fukuda, Y., \EtAl\ 1998,
  Phys.\ Rev.\ Lett.,\  81, 1562 
  
\bibitem[Gonzalez-Garcia \EtAl(2000)]{Gonzalez-Garcia:2000aj}
  Gonzalez-Garcia, M.~C., de Holanda, P.~C., Pena-Garay, C. \& Valle,
  J.~W. 2000,
  Nucl.\ Phys.\ B, 573, 3
  
\bibitem[Guzzo \EtAl(2001)]{Guzzo:2001mi}
  Guzzo, M., \EtAl\ 2001,
  Nucl.\ Phys.\ B, 629, 479 
  
\bibitem[Hall, Kostelecky, \& Raby(1986)]{hal86}
  Hall, L.~J., Kostelecky, V.~A., \& Raby, S. 1986,
  Nucl.\ Phys.\ B, 267, 415
  
\bibitem[Kirsten(2002)]{gallex}  
  Kirsten, T.  2002, Talk at  Neutrino 2002,
  \verb"http://neutrino2002.ph.tum.de"
  
\bibitem[Lim \& Marciano(1988)]{lim88}
  Lim. C.~S., \& Marciano, W.~J. 1988,
  Phys.\ Rev.\ D, 37, 136
  
\bibitem[Maltoni \EtAl(2002a)]{Maltoni:2002ni} 
  Maltoni, M., Schwetz, T., Tortola, M.~A. \& Valle, J.~W. 2002a, 
  Phys.\ Rev.\ D {\bf 67}, 013011, preprint hep-ph/0207227.
  
\bibitem[Maltoni \EtAl(2002b)]{Maltoni:2002aw}
  Maltoni, M., Schwetz, T., \& Valle, J.~W.~ 2002b, preprint (hep-ph/0212129)
  
\bibitem[Mestel \& Weiss(1987)]{MestelWeiss} 
  Mestel, L., \& Weiss, N.~O. 1987, MNRAS, 
  226, 123
  
\bibitem[Mikheev \& Smirnov(1985)]{mik85}
  Mikheev, S.~P., \& Smirnov, A.~Y. 1985,
  Sov.\ J.\ Nucl.\ Phys.,\  42, 913
  
\bibitem[Miranda \EtAl(2001a)]{mir01a} 
  Miranda, O.~G., \EtAl\ 2001a,
  Nucl.\ Phys.\ B, 595, 360 
  
\bibitem[Miranda \EtAl(2001b)]{mir01b} 
  Miranda, O.~G., \EtAl\ 2001b,
  Phys.\ Lett.\ B, 521, 299
  
\bibitem[Mohapatra \& Valle(1986)]{moh86} 
  Mohapatra, R.~N., \& Valle, J.~W.~F. 1986
  Phys.\ Rev.\ D, 34, 1642
  
\bibitem[Nunokawa \EtAl(1996)]{Nunokawa:1996qu}
  Nunokawa, H., Rossi, A., Semikoz, V.~B., \& Valle, J.~W. 1996
  Nucl.\ Phys.\ B, 472, 495
  
\bibitem[Parker(1979)]{Parker}
  Parker, E.~N. 1979, Cosmical Magnetic Fields, Clarendon Press, Oxford
  
\bibitem[Schechter \& Valle(1980)]{Schechter:1980gr}
  Schechter, J., \&  Valle, J.~W. 1980,
  Phys.\ Rev.\ D,  22, 2227
  
\bibitem[Schechter \& Valle(1981)]{Schechter:1981hw}
  Schechter, J., \& Valle, J.~W. 1981,
  Phys.\ Rev.\ D,  24, 1883 
  [Erratum-ibid.\ D  25, 283 (1982)]
  
\bibitem[Wolfenstein(1977)]{Wolfenstein:1977ue}
  Wolfenstein, L. 1977
  Phys.\ Rev.\ D, 17, 2369 
  
\end{thebibliography}
\end{document}